\documentclass[aps,twocolumn,showpacs,preprintnumbers,amsmath,amssymb,superscriptaddress]{revtex4}

\usepackage{graphicx}
\usepackage{color}
\usepackage{dcolumn}
\usepackage{bm}

 \usepackage{graphicx}
\usepackage{amsmath}

\begin{document}

\title{Boolean modeling of collective effects in complex networks}

\author{Johannes Norrell}
\author{Joshua E.~S.~Socolar}
\affiliation{Center for Nonlinear and Complex Systems, IGSP Center for Systems Biology, and Physics Department, Duke University, Durham, NC, U.S.A.}

\date{\today}

\begin{abstract}
Complex systems are often modeled as Boolean networks in attempts to capture their logical structure and reveal its dynamical consequences. Approximating the dynamics of continuous variables by discrete values and Boolean logic gates may, however, introduce dynamical possibilities that are not accessible to the original system.  We show that large random networks of variables coupled through continuous transfer functions often fail to exhibit the complex dynamics of corresponding Boolean models in the disordered (chaotic) regime, even when each individual function appears to be a good candidate for Boolean idealization.  A suitably modified Boolean theory explains the behavior of systems in which information does not propagate faithfully down certain chains of nodes.  Model networks incorporating calculated or directly measured transfer functions reported in the literature on transcriptional regulation of genes are described by the modified theory.
\end{abstract}




\maketitle

\section{Introduction}

Natural systems often involve many types of elements interacting in a complicated fashion.  The interactions may be difficult to describe, and may be mediated in ways that are poorly understood.  In this situation, it is necessary to find a model that captures most of the salient features of the system without attempting to describe all the details.

Boolean networks are often constructed to model the logic of systems with a complex set of interactions \cite{kauffman1,kauffman2,aldana}.  In this idealization, continuous variables are modeled by binary states, and interactions are modeled by Boolean update rules.  A binary representation is a natural approximation for systems whose elements tend to take distinct high and low values with sharp transitions between states.  However, even when individual elements are good candidates for Boolean modeling, qualitative discrepancies between the dynamics of the underlying system and its Boolean idealization can arise.

The relation between continuous and Boolean systems has been the subject of some study.  In previous work, we identified the features that cause a discrepancy in the attractor dynamics of small systems \cite{norrell}.  Davidich and Bornholdt have shown that through a careful examination of the attractor dynamics, a Boolean model can be constructed that faithfully reproduces the temporal sequence of states obtained from direct binarization of a given continuous system. \cite{davidich}.  Glass et.\ al.\ have extensively studied a class of large networks governed by piecewise linear differential equations, which involve both Boolean and continuous variables.  They have noted that artifacts introduced by synchronous update contribute significantly to the size of the attractor set, and that both periodic and chaotic dynamics can be observed, although chaos is quite uncommon in random networks with a connectivity of $k=2$ \cite{glass1,glass2}.  Magnasco has shown that it is possible to construct continuous systems that implement any specified Boolean computation \cite{magnasco}.  The present work addresses new dynamical features that arise in large networks of generic elements with sigmoidal response functions and explores the extent to which they can be understood using Boolean models.

We consider an illustrative class of continuous models and show that information propagation along chains plays a key role in determining the qualitative dynamical behavior of large random networks. We develop a modified Boolean theory incorporating the effects of signal decay on certain chains that explains key features of the continuous dynamics. Applying the theory to the well-known transition between order and disorder in random networks \cite{samuelsson,kesseli} reveals that signal decay has little effect on ordered dyanmics but can lead to a substantial suppression of disorder. Finally, we study cis-regulatory functions from the quantitative biology literature and show that our theory accounts well for the dynamics of random networks constructed using those functions. The differences in collective behavior of the continuous systems and naive Boolean models are not simple extensions of the differences in attractor structure noted in earlier studies \cite{norrell}.  They suggest both that caution should be taken when making inferences based on Boolean modeling of individual nodes and that appropriately modified Boolean models can still provide useful insights.

\section{A class of continuous models}

We study continuous systems with variables $\{x_i\}$ and time evolution equations
\begin{align}\label{eqn:transferfuncs}
\dot{x_i}(t)&=f_i[x_j(t-\tau_{ij}),x_k(t-\tau_{ik})]-\gamma_ix_i(t)~,\\
f_i(x_j,x_k)&=\eta_i\left[\frac{1+d_j x_j^2+d_k x_k^2+d_{jk} x_jx_k}{1+b_j x_j^2+b_k x_k^2+b_{jk} x_jx_k}\right]~,
\end{align}
where $\eta_i$, $\gamma_i$, the $b$'s and the $d$'s are constant coefficients and $\tau_{ij}$ is a constant time delay associated with the conversion of the output of node $j$ into its active form and/or a propagation time between node $j$ and its target node $i$.  Each node $i$ receives exactly two inputs from randomly selected nodes in the network and responds as determined by its transfer function $f_i$.  The form of the differential equations is motivated by studies of genetic regulatory networks, in which case the variables represent mRNA concentration \cite{andrecut}.

\begin{figure}
\includegraphics[width=.48\textwidth]{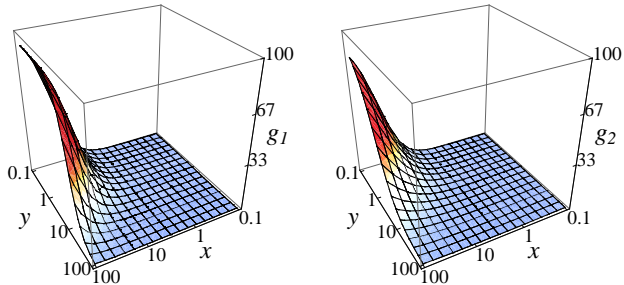}
\caption{\label{fig:g} The transfer functions $g_1(x,y)$ (left) and $g_2(x,y)$ (right).  Both functions have the same Boolean idealization.}
\end{figure}

\begin{table}[b]
\caption{\label{table:params} Parameter values for transfer functions.}
\vspace{1mm}
\begin{tabular}{c | ccccccc}
\hline
Function  & $~d_j~$ & ~$d_k~$ & $~d_{jk}~$ & $~b_j~$ & $~b_k~$ & $~b_{jk}~$ \\
\hline
$g_1$ & .1 & 0 & 0 & .001 & .1 & 0 \\
$g_2$ & .1 & 0 & 0 & .001 & 0 & .1\\
$h$  & .05 & .05 & 0 & .0005 & .0005 & 0 \\
\hline
\end{tabular}
\end{table}

Time delays are included for two complementary reasons.  First, the interactions between elements in the systems of interest typically involve a series of events that take time to complete.  For transcriptional networks, the activation or repression of a target gene due to the buildup of a particular mRNA requires the translation of the mRNA, the folding of the protein, and the transport of the protein to the nucleus.  In this case, the delays in our model equations represent the time required for protein levels to build up, along with whatever other post-translational processes are required and may be thought of as capturing the dominant effect of a set of explicit equations for additional variables.  Second, the delays are necessary to produce simple oscillatory behavior in small negative feedback loops.   A self-repressor in a Boolean system produces oscillations rather than fixed points,  but a continuous self-repressor will not oscillate in the absence of a time delay.  To get a meaningful comparison of large Boolean network models and the underlying continuous systems, we want to study continuous models capable of exhibiting the oscillatory behaviors of very simple negative feedback loops. Though there may be cases in which the function of negative feedback in a real system is just to regulate the level of a fixed point, such feedback would be irrelevant from the perspective of Boolean modeling.

A system that nicely illustrates the role of time delays is the repressilator (a loop of three repressors) studied by Elowitz and Leibler \cite{elowitz}.  To model the observed oscillations, it was necessary to include separate equations for the net rate of production of RNA and proteins.  Coupled equations for RNA concentrations of the form of Eq.~\ref{eqn:transferfuncs} with $\tau=0$ yield only a stable fixed point \cite{elowitz}.  With sufficiently large time delays, however, the fixed point becomes unstable to an oscillatory attractor; it is not necessary to include explicit equations for protein production.

\section{Analysis of two specific systems}

We consider two random network systems, $S_1$ and $S_2$, that have the same Boolean idealization but turn out to behave quite differently.  Each node in $S_\alpha$ is assigned $f_i=g_\alpha$ with probability $p$ and $f_i=h$ otherwise, where $g_\alpha$ and $h$ are defined by the parameters listed in Table~\ref{table:params}.  $g_1$ and $g_2$ are plotted in Fig.~\ref{fig:g}.  Both would be approximated by a Boolean {\sc nif} function, which returns a 1 if and only if the first input is 1 and the second input is 0.  $h$ corresponds to a Boolean {\sc or} function.  The choice of  {\sc nif} and {\sc or} functions allows  us to tune through the order--disorder transition by varying $p$.   The qualitative results are not specific to this choice. 

The continuous dynamics are simulated using the fourth order Runge-Kutta method described in \cite{norrell}.  The time delays $\{\tau_{ij}\}$ are set to 1, the decay constants $\{\gamma_i\}$ are chosen at random from the interval [0.8,1.2], and the normalization constants $\{\eta_i\}$ are chosen such that $g_1$, $g_2$ and $h$ all have the same saturation value of 100.  The results presented here are not sensitive to the values of  $\{\tau_{ij}\}$ or $\{\gamma_i\}$.  We studied networks of size $N=1000$.  For each given distribution of transfer functions, we simulated between 15 and 30 networks, with 15 to 30 random initial conditions each, each network requiring about one hour of computation time on a desktop computer.

\begin{figure}[t]
\includegraphics[width=.48\textwidth]{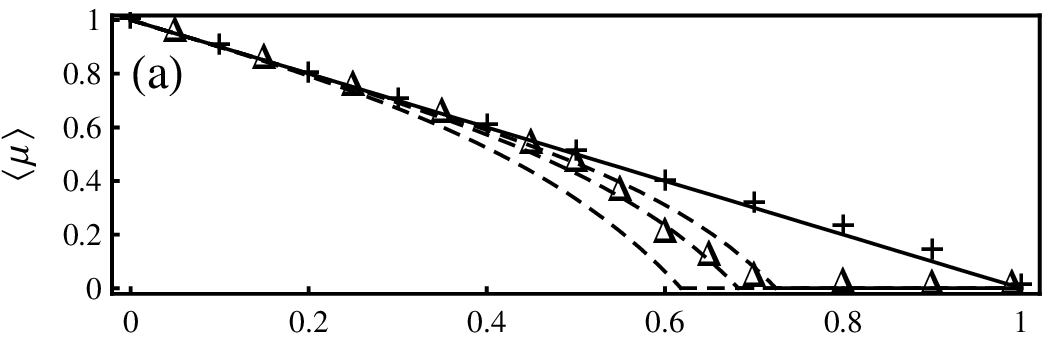}
\includegraphics[width=.48\textwidth]{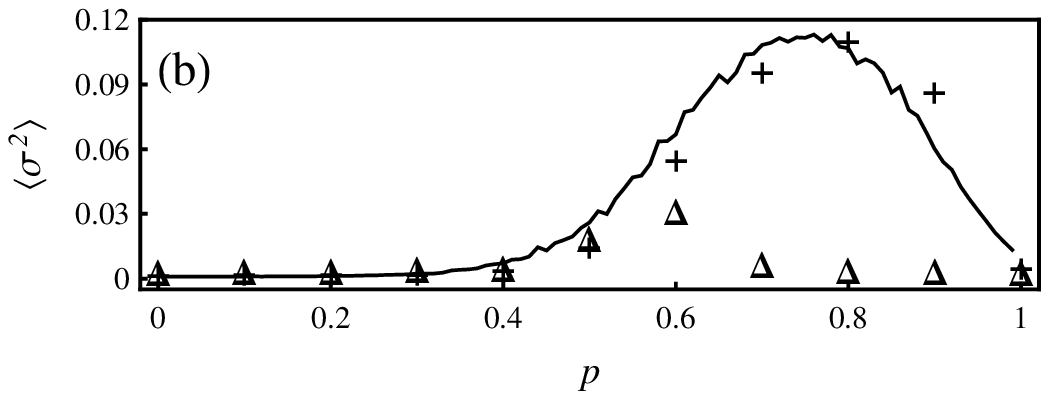}
\caption{\label{fig:meanvar} 
(a) The mean fraction of {\sc on} nodes.  The $+$ and $\Delta$ symbols correspond to $S_1$ and $S_2$, respectively.  The solid line $\mu=1-p$ is the prediction for an infinite random Boolean network with a fraction $p$ of {\sc nif} gates and $1-p$ of {\sc or} gates. The dashed curves show the theoretical results discussed in the text.  (b) The total variance corresponding to the systems in (a). The solid curve is the average variance from simulations of the Boolean model.}
\end{figure}

To compare the dynamics of the continuous systems with their Boolean counterparts, we binarize the continuous time series, setting a node's value to 1 (0) if it is above (below) a specified threshold. For a given network, let $\varphi_i$ be the mean value of node $i$ over time, obtained after concatenating time series of equal duration from attractors generated by different initial conditions.  We focus on two quantities: $\mu={\rm Avg}[{\varphi_i}]$, and $\sigma^2={\rm Avg}\left[\varphi_i-\varphi_i^2\right]$, where ${\rm Avg}\left[\ldots\right]$ denotes an average over nodes.  $\sigma^2$ is the average of the variances for a system with binary values 0 and 1.  Note that frozen nodes, which go to the same static value at long times on all attractors, contribute zero to $\sigma^2$.   The concatenation of attractors allows nodes that are constant on any single attractor, but not at the same value on all attractors, to contribute to $\sigma^2$.

Fig.~\ref{fig:meanvar} shows the ensemble averages $\left<\mu\right>$ and $\left<\sigma^2\right>$ as functions of $p$.  $S_1$ behaves very much like its Boolean idealization, represented by the solid curves; $S_2$ does not.  We use established techniques \cite{moreira, shmulevich,kesseli,ramo} to determine that the Boolean system is ordered for $p\in[0,0.5)$, disordered for $p\in(0.5,1)$ and critical for $p=0.5$ and $p=1$.  In the large system limit, we expect $\left<\sigma^2\right>=0$ in the ordered regime and a continuous transition to $\left<\sigma^2\right>>0$ in the disordered regime.  The nonzero value of $\left<\sigma^2\right>$ at the critical points is a finite size effect.  The variance of $S_2$ is strongly suppressed in the region where the Boolean model is disordered.  We note that $\left<\mu\right>=0$ implies $\left<\sigma^2\right>=0$, so for this particular system, we can explain the suppression of $\left<\sigma^2\right>$ by deriving a theory of $\left<\mu\right>$.  The dashed curves explaining the behavior of $S_2$ are based on the theory discussed below.

\begin{figure}[t]
\includegraphics[width=.48\textwidth]{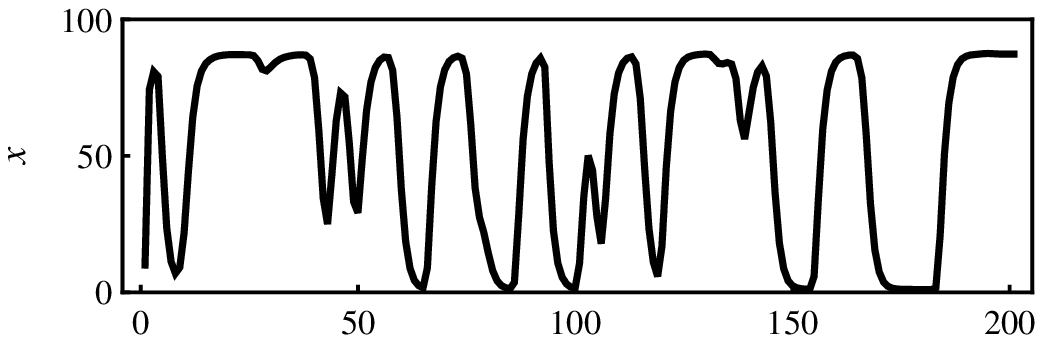}
\includegraphics[width=.48\textwidth]{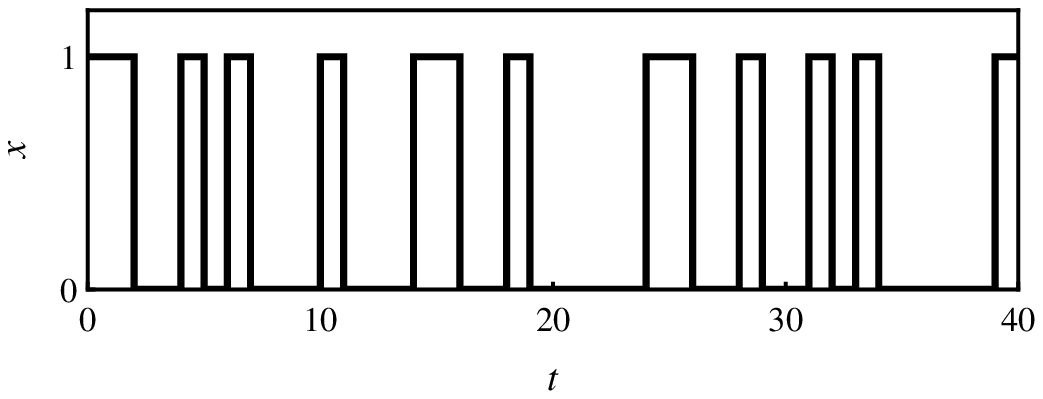}
\caption{\label{fig:series} 
Typical time series for an unfrozen node in the disordered regime ($p = 0.8$) in a continuous system, $S_1$, and in a synchronously updated Boolean system.  Both systems have attractors with complex dynamics.}
\end{figure}

As discussed by Magnasco in \cite{magnasco}, a system whose transfer functions are globally compatible is capable of executing logical operations, where global compatibility means that all transfer functions have the same two fixed point values; i.e., that when every input is held steady at either the high or low value, every output will also take one of those two values.  We show here a system ($S_1$) that behaves statistically like its Boolean analogue, although the transfer functions for the two node types are not tuned to be globally compatible and the attractors are not always steady states.  Fig.~\ref{fig:series} shows a typical time series for a continuous node in $S_1$ and a Boolean node in networks with $p=0.8$. Both exhibit the complicated behavior associated with the disordered regime.

Certain features of continuous systems are known to strongly affect the number and nature of attractors in small systems and simple rings in ways that are not captured by Boolean models \cite{norrell}.   It is not clear, however, whether these features lead to important effects in large systems.  The close match for $\left<\mu\right>$ and $\left<\sigma^2\right>$ of $S_1$ and its Boolean model (Fig.~\ref{fig:meanvar}) suggests that they do not, but care must be taken to interpret these results.  Much of the agreement can be attributed to the fact that $\left<\mu\right>$ and $\left<\sigma^2\right>$ are determined primarily by the fraction of frozen nodes in the network, a quantity that does not depend on the timing or sequence of updating the nodes \cite{samuelsson}.

In the ordered regime, the nearly vanishing number of nodes that are {\it not} frozen leads to very low values of $\left<\sigma^2\right>$.   The agreement in the disordered regime is less trivial.  Consider, for example, a ring of nodes containing a random mixture of copiers and inverters.  For synchronously updated Boolean dynamics, every attractor of a given network has a partner in which the values 0 and 1 are exchanged for all time steps, and these two attractors have basins of the same size.  (When the number of inverters is odd, every attractor is its own partner.)  Thus $\mu$ is always $1/2$ and $\sigma^2$ is always the maximum value of $1/4$.  As discussed in \cite{norrell}, the breaking of on-off symmetry in generic continuous systems leads to a collapse of almost all attractors, leaving only two fixed point states for rings with an even number of inverters and one oscillating state for rings with an odd number of inverters.  The odd case produces $\left<\mu\right>\approx 0.5$ and $\left<\sigma^2\right>\approx 0.25$.  These are not strict equalities  because, unlike for the synchronous Boolean ring, a node in the corresponding continuous system need not spend precisely the same time in the high and low states, but the symmetry breaking effect is typically small. \cite{norrell}.  In the even case, on the other hand, the two fixed point states can have dramatically different size basins and therefore exhibit a substantially reduced value of $\sigma^2$.  For a ring of size $N=100$ with a random mixture of continuous copiers and inverters, we find $\left<\mu\right>\approx0.5$ and $\left<\sigma^2\right>\approx0.12$.  So in the case of simple rings, where $\left<\mu\right>$ and $\left<\sigma^2\right>$ are determined by dynamical properties of the attractors rather than numbers of frozen nodes, we do not see agreement similar to that of  $S_1$ and the Boolean system shown in Fig.~\ref{fig:meanvar}.  It appears that the complex network of connections between active rings in the disordered regime restores the agreement.  Fig.~\ref{fig:series} suggests that Boolean and continuous attractors in the disordered regime have similar temporal features, but a full characterization of the dynamics of individual attractors is beyond our present scope.

\begin{figure}
\includegraphics[width=.48\textwidth]{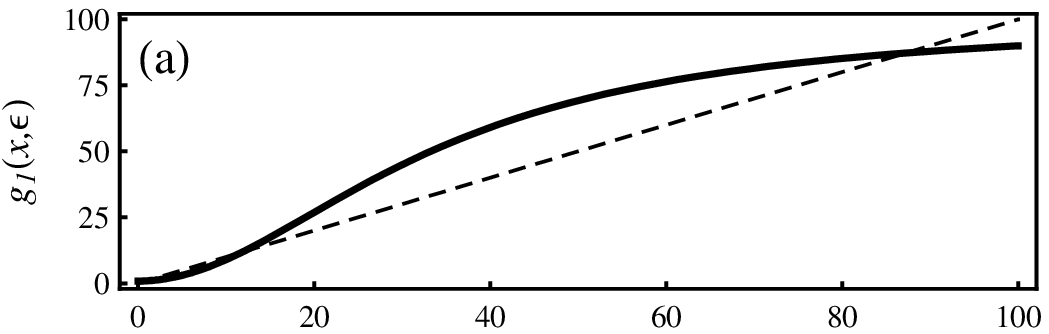}
\includegraphics[width=.48\textwidth]{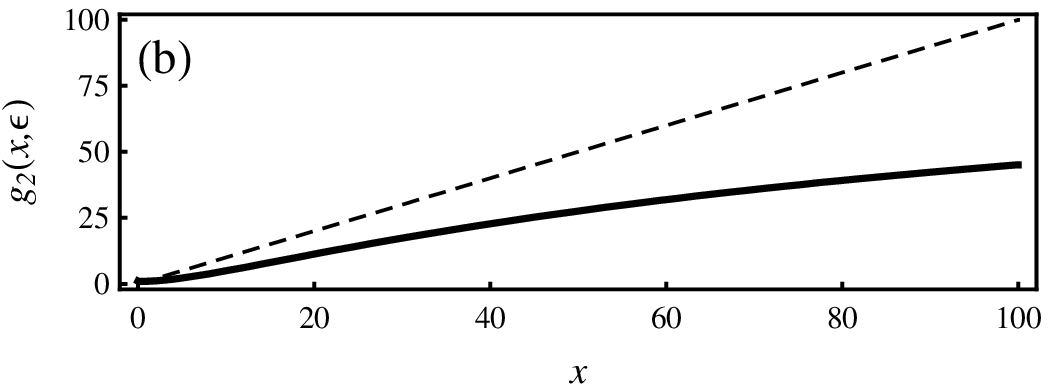}
\caption{\label{fig:nifcopy} 
(a) $g_1(x,\epsilon)$ and (b) $g_2(x,\epsilon)$, with $\epsilon=1.12$.}
\end{figure}

We now turn to the analysis of $S_2$,  a case where the continuous dynamics show substantial deviations from the naive Boolean expectations.  To understand why $S_1$ and $S_2$ behave differently, we examine the functions $g_1(x,y)$ and $g_2(x,y)$ when the second input is held at a low value $\epsilon=1.12$, the low stable fixed point of $h(x,x)$.  A Boolean {\sc nif} function whose second input is held at 0 acts as a copier on its first input, so we expect the functions $g_1(x,\epsilon)$ and $g_2(x,\epsilon)$ to behave like copiers.  As shown in Fig.~\ref{fig:nifcopy}, $g_1(x,\epsilon)$ has two stable fixed points, but $g_2(x,\epsilon)$ has only one stable fixed point.

As noted in \cite{magnasco}, the faithful propagation of information along chains of nodes requires two stable fixed points in the transfer function.  We refer to the loss of information along chains of nodes lacking a second stable fixed point as ``propagation failure.''  Because $g_2$ does not have the required fixed point structure, propagation failure may cause the system $S_2$ to behave differently from its naive Boolean idealization.  We now present a modified Boolean model that accounts for propagation failure and agrees well with simulations, as shown by the dashed curves in Fig.~\ref{fig:meanvar}.  The success of this theory indicates that for large random networks, propagation failure is the primary source of the measured discrepancy between the continuous system and the original Boolean model.

A modified Boolean theory for $S_2$ is derived by noting that some of the nominal {\sc nif} nodes are actually better modeled by the Boolean {\sc off} function.  Consider a chain of continuous nodes where each has the transfer function $g_2(x,y)$, and where node $i$ is the first input ($x$) into node $i+1$.  Because $g_2$ lacks the necessary high fixed point, propagation failure prevents nodes far down the chain from ever rising above threshold in response to a high input signal to the first node in the chain ($i=1$).  Though they may be initialized at a high value, they will always stay low after a brief transient.  Let node $m$ be the first node in the chain that cannot rise above threshold due to propagation failure.  We will model node $i$ with a {\sc nif} function if $i<m$, and with an {\sc off} function if $i\geq m$.  The value of $m$ for the a particular chain of $g_2$ nodes will depend on the parameters in $g_2$, the choice of threshold, and the value of the high input into the first node of the chain.  Approximating the system as a random Boolean network with fractions $r$, $q$, and $1-r-q$ of {\sc off}, {\sc nif}, and {\sc or} nodes, respectively (neglecting correlations in the positions of the {\sc off} nodes), $\left<\mu\right>$ can be calculated exactly as the stable fixed point of the bias map \cite{ramo}

\begin{align}
\label{eq:biasmap} \rho_{t+1}=q\left(\rho_t-\rho_t^2\right)+(1-r-q)\left(2\rho_t-\rho_t^2\right)~,
\end{align}
where $\rho_t$ represents the average fraction of nodes with a value of 1 at time $t$.  The stable fixed point is
\begin{align}
\label{eq:mu1} \left<\mu\right>&=\max\left\{0,\frac{1-2r-q}{1-r}\right\}\,.
\end{align}

Let $n$ be the value of $m$ for $S_2$, computed using the average value of the high input signals that arise from the dynamics.  For a large random network, the fraction of nodes that are assigned the {\sc off} function is $r=p^n$, so the fraction of nodes that truly act as {\sc nif} is $q=p-p^n$.  Substitution into Eq.~\ref{eq:mu1} gives
\begin{align}
\label{eq:mu2} \left<\mu\right>&=\max\left\{0,\frac{1-p-p^n}{1-p^n} \right\}\,.
\end{align}
The prescription for finding {\sc off} nodes does not explicitly account for the failure of propagation around rings of $g_2$ nodes smaller than $n$, but such loops are very rare in random networks.

The dashed curves in Fig.~\ref{fig:meanvar} show $\left<\mu\right>$ for $n\in\{2,3,4\}$, with $n$ increasing to the right .  The simulations suggest a crossover from $n=4$ for $p\lesssim0.5$ to $n=3$ for $p\gtrsim0.5$.  The reduction of $n$ arises because the average value taken by nodes that are above threshold decreases as $p$ increases, which implies lower input values to chains of $g_2$ nodes.  The switch from $n=4$ to $n=3$ associated with the function $g_2(x,\epsilon)$ of Fig.~\ref{fig:nifcopy} and our chosen threshold of 10 occurs when the initial input to a chain is about 58.  Simulations reveal that the average value of nodes above threshold crosses 58 at $p\approx0.51$, which corresponds reasonably well with the crossover observable in Fig.~\ref{fig:meanvar}.

The suppression of disorder, as indicated by small values of $\left<\sigma^2\right>$ for $p>0.5$ in Fig.~\ref{fig:meanvar}(b), is caused by the effective insensitivity of nodes at the end of sufficiently long chains.  Because nodes in this set have approximately fixed values, all nodes receiving both inputs from this set will also have approximately fixed values, leading to a cascade of effectively frozen nodes.  The net result is a substantial loss of sensitivity for the network as a whole.

\begin{figure}[t]
$\begin{array}{cc}
\includegraphics[scale=0.4]{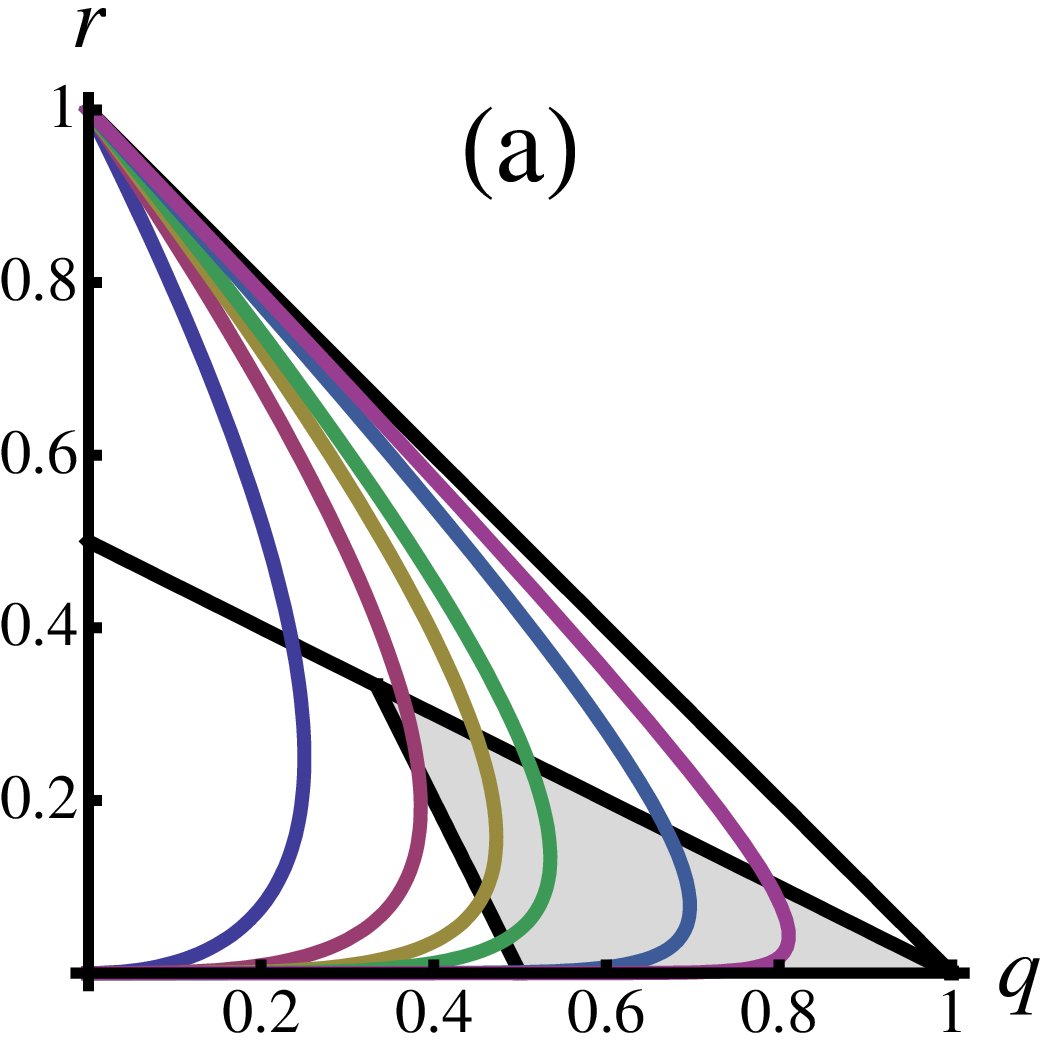}&\includegraphics[scale=0.4]{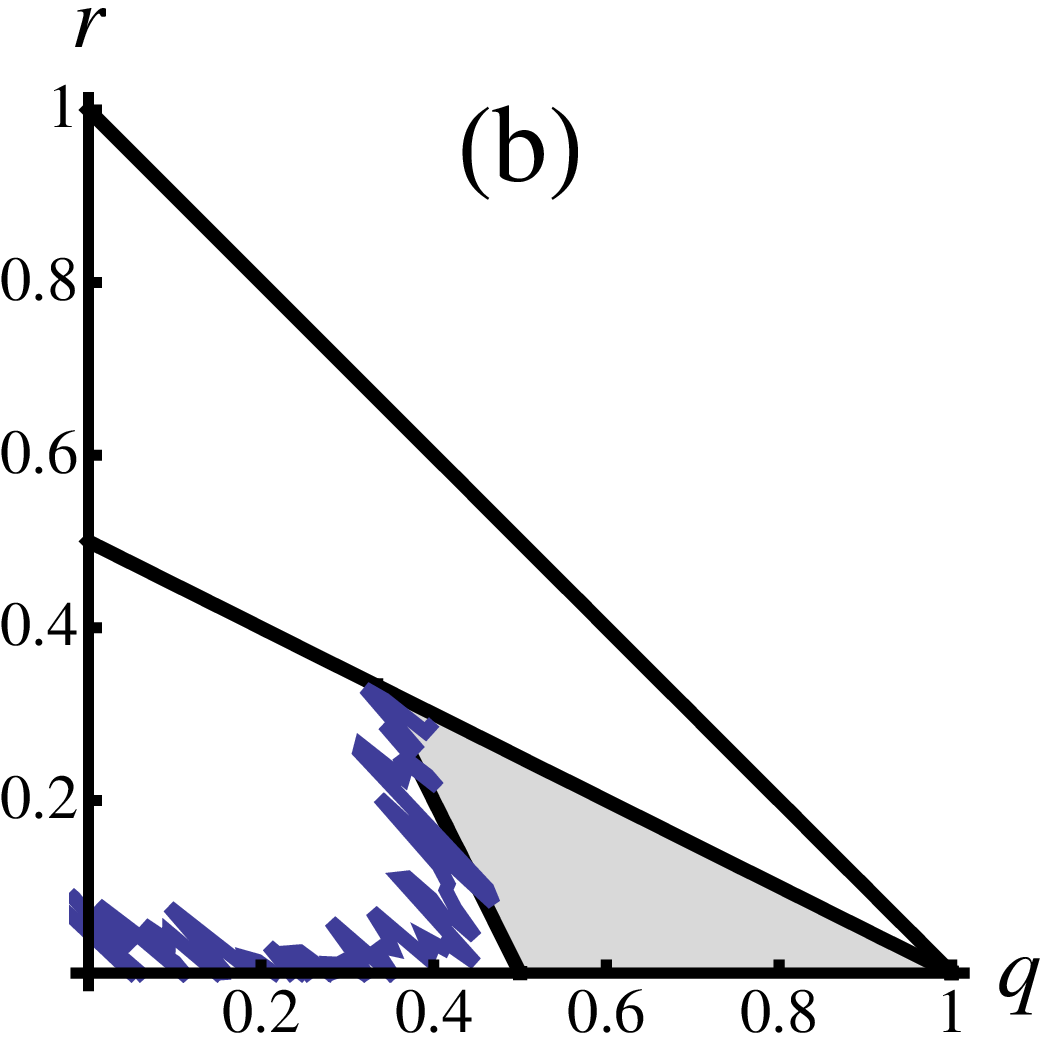}
\end{array}$
\caption{\label{fig:qrplane} 
(a) Trajectory in the $q-r$ plane, as predicted by the mean field model, for fixed decay lengths $n\in\{2,3,4,5,10,20\}$.  The curves go from low to high $n$ when viewed left to right.  (b) Fit of actual data with varying $n$.  See text for details.}
\end{figure}

Further insight into the order--disorder transition comes from examining the trajectory traced in the $q-r$ plane as $p$ increases from 0 to 1, shown in Fig.~\ref{fig:qrplane}.  The shaded sector is the disordered regime of the Boolean system, where the slope of the Derrida plot exceeds unity \cite{derrida,kesseli}.  The unshaded sectors correspond to ordered regimes, and the boundaries to critical systems. Fig.~\ref{fig:qrplane}(a) shows trajectories corresponding to different $n$.  We can measure $\left<\mu\right>$ in simulations of $S_2$ and determine $r$ and $q$ from the relations $r+q=p$ and Eq.~\ref{eq:mu1}. The results are shown in Fig.~\ref{fig:qrplane}(b).  Only the results for $\left<\mu\right>>0$ are plotted; the theory cannot determine unique values for $r$ and $q$ in the upper unshaded sector because every $(q,r)$ pair gives $\left<\mu\right>=0$.  This plot again reveals a shift from $n=4$ to $n=3$, but also shows that the system skirts a critical boundary as $p$ is varied.  The dynamical suppression of disorder in this case provides a mechanism for keeping a network in the critical regime over a wide range of parameter values.  We note, however, that this is not necessarily a generic effect.  Functions producing larger values of $n$ would permit some degree of disorder.

\section{Results for reported transfer functions}

It is instructive to examine suggested or measured transfer functions from the literature to see whether collections of similar functions would faithfully execute their nominal Boolean logic.  In \cite{hwa}, Boltzmann weights are used to compute the probability that RNA-polymerase will bind to the promoter region of a gene as a function of the concentrations of the transcription factors for that gene.  If we take the transfer functions to be proportional to the probability functions, we find that they lack the necessary fixed points for propagating information and that random networks built from them exhibit strong suppression of disorder.  We note, however, that there is evidence for cooperative effects and post-transcriptional processes that influence the effective transfer function, so we do not necessarily expect the systems considered below to be representative of real regulatory networks.

\begin{figure}[tbp]
\includegraphics[width=.48\textwidth]{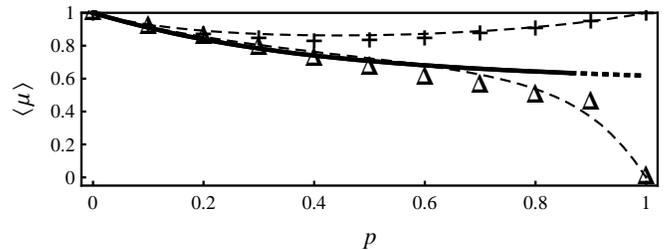}
\caption{\label{fig:hwa}
Mean node value of a network using {\sc nand} and {\sc or} transfer functions from \cite{hwa} and a threshold that is ($+$ symbols) below, and ($\Delta$ symbols) above the intermediate fixed point defined in the text.  $p$ is the fraction of {\sc nand} nodes.  Our mean field theory (dashed curves) gives a much better approximation of the dynamics than does the naive Boolean model (solid curve).}
\end{figure}

Consider a continuous system with two transfer functions defined in \cite{hwa} as implementing {\sc nand} and {\sc or} logic.  A mean field theory analogous to Eq.~\ref{eq:mu2} can be constructed.  The results in this case depend upon the choice of threshold for the binarization of the time series.  Two plots for $\left<\mu\right>$, using different threshold values, are given in Fig.~\ref{fig:hwa}.  The solid curve shows the behavior of the straightforward Boolean idealizations, and the dashed curves show the predictions made by our mean field theory.  The theory predicts $\left<\sigma^2\right>=0$ for all fractions $p$ of {\sc nand} nodes because propagation failure pushes the system into the ordered phase, an effect confirmed by simulations.  As above, accounting for the insensitivity of some nodes allows a reasonably accurate prediction of the dynamics of the continuous system.

In a random Boolean network of {\sc nand} and {\sc or} gates, the fixed point in the bias map becomes unstable to a 2-cycle at $p=\sqrt{3}/2$.  Oscillations in bias occur also in continuous systems with long time delays and short decay times, but not if these time scales are comparable.  In the latter case, the {\sc nand} nodes find stable intermediate values.  Fig.~\ref{fig:hwa} shows the dynamics with the choice $\tau_{ij}=1$ and $\gamma_i\in[0.8,1.2]$ with two different threshold values, one above and one below the intermediate fixed point of the all {\sc nand} system.  Analysis of continuous systems with oscillating bias is beyond the scope of this work.

Ref.~\cite{hwa} provides no obvious guide for choosing the normalization constants for the transfer functions.  For Fig.~\ref{fig:hwa} we used $\eta_1=\eta_2=1000$. We tried every combination of $(\eta_1,\eta_2)$ with $\eta_i\in\{200,400,600,800,1000\}$ and found no pair that produced behavior significantly different from those shown.  In this case, as in most cases, the {\sc nand} nodes took clearly separated high and low values for most values of $p$, but the {\sc or} nodes typically converged to a single value due to the absence of a low fixed point.

A second example of a biological transfer function is the {\it lac} cis-regulatory input function studied in \cite{setty}.  The {\it lac} promoter requires both IPTG and cAMP to be produced.  Setty et.\ al.\ discuss both the real input function and an idealized continuous {\sc and} function.  Neither of these functions satisfies the criterion for faithful information propagation in a network.  For high values of the first input, the function does not have a stable high fixed point for the second input. A large network of such elements would not express its nominal Boolean logic.

It has been shown both theoretically and experimentally that signaling cascades can create switch-like responses \cite{ferrell,kholodenko, hooshangi} in genes several steps downstream from the initial stimulus.  In transcriptional regulation, one mechanism for achieving the required characteristics of the transfer functions involved was proposed in \cite{hermsen}, wherein an evolutionary model generated steeper transfer functions by introducing auxiliary binding sites for a transcription factor.  Another mechanism has been proposed in recent work by Buchler et.\ al., who emphasize the possible role of protein or RNA sequestration in creating sharp transfer functions \cite{buchler1,buchler2}.  We note here that the demonstration of an effective Hill function with high cooperativity does not by itself imply that a cascade of similar elements would successfully propagate information.  One needs to know how to normalize the output levels to see whether there is a high fixed point, and this requires knowledge of the level needed to activate the downstream target.  Nevertheless, the use of cooperativity  and sequestration to construct elements with nearly Boolean behavior appears promising.  Our present work highlights another relevant issue for synthetic or natural transcriptional circuits: the interactions of transcription factors may make it difficult to simultaneously meet all of the requirements for signal propagation for genes with two or more inputs.  Attempts to experimentally demonstrate a sharp, two-input logic gate may yield new insights into the feasibility of implementing complex logic in biological systems.

It is not clear, however, that biological networks need to be able to carry out arbitrarily complicated logical operations. Our results show that faithful propagation down long chains is not crucial for implementing the Boolean dynamics arising in the ordered and critical regimes in systems of a few thousand elements.  Moreover, if operation near criticality is advantageous for a biological system, as suggested in \cite{kauffman1}, suppression of disorder may be beneficial.  We have identified a dynamical mechanism for accomplishing this, which may be a useful alternative to alteration of the network architecture.  A limited range of propagation could actually be an important feature of real biological networks.

\section{Conclusion}

In conclusion, we have seen that the detailed form of continuous transfer functions can have a qualitative effect on the dynamics of large random networks.  If all of the transfer functions have suitable fixed point structures such that signals can propagate down chains of arbitrary length, fundamental statistics of the attractors are similar.  This indicates that disordered systems are not prone to the type of attractor collapse that occurs on simple rings.  When some of the transfer functions do not have a suitable fixed point structure, the dynamics in the disordered regime are strongly affected. In the case studied above, the suppression of disorder leads to an extended domain in the parameter space where the system is close to a critical Boolean network. A successful theory of these effects has been obtained within a Boolean framework by accounting for the failure of information to propagate down long chains of nodes. Further characterization of the temporal structure of individual attractors in the disordered regime is needed to fully understand the importance of these phenomena in large networks.

{\it Acknowledgments} -- We would like to thank S.A. Kauffman and M.
Andrecut for providing motivation and inspiration.  We would also like
to thank V. Sevim, X. Gong and X. Cheng for helpful discussions.  This
work was supported by grants from NSF (PHY-0417372) and NIH
(1P50-GM081883).


\begin{thebibliography}{100}
\bibitem{kauffman1}S.A. Kauffman.  {\it The Origins of Order: Self-Organization and Selection in Evolution}.  Oxford University Press (1993).
\bibitem{kauffman2}S.A. Kauffman.  Metabolic stability and epigenesis in randomly constructed genetic nets, {\it J. Theor. Biol.} {\bf 22}, 437 (1969).
\bibitem{aldana}M. Aldana, S. Coppersmith and L.P. Kadanoff.  Boolean dynamics with random couplings, {\it Springer Applied Mathematical Sciences Series} {\bf 23} (2003).
\bibitem{norrell}J. Norrell, B. Samuelsson and J.E.S. Socolar.  Attractors in continuous Boolean network {\it Phys. Rev. E} {\bf 76}, 046122 (2007).
\bibitem{davidich}M. Davidich and S. Bornholdt.  From differential equations to Boolean networks: a case study in modeling regulatory networks, arXiv:0807.1013v2 (2008).
\bibitem{glass1}R.J. Bagley and L. Glass.  Counting and classifying attractors in high dimensional dynamical systems, {\it J. Theor. Biol.} {\bf 183}, 269 (1996).
\bibitem{glass2}L. Glass and C. Hill.  Ordered and disordered dynamics in random networks, {\it Europhys. Lett.} {\bf 41}, 599 (1998).
\bibitem{magnasco}M.O. Magnasco.  Chemical kinetics is Turing universal, {\it Phys. Rev. Lett.} {\bf 78}, 1190 (1997).
\bibitem{samuelsson}B. Samuelsson and J.E.S. Socolar.  Exhaustive percolation on random networks, {\it Phys. Rev. E} {\bf 74}, 036113 (2006).
\bibitem{kesseli}J. Kesseli, P. R\"am\"o and O. Yli-Harja.  Iterated maps for annealed Boolean networks, {\it Phys. Rev. E} {\bf 74}, 046104 (2006).
\bibitem{andrecut}M. Andrecut and S.A. Kauffman.  Mean-field model of genetic regulatory networks, {\it New J. Phys.} \textbf{8}, 148 (2006).
\bibitem{elowitz}M.B. Elowitz and S. Leibler.  A synthetic oscillatory network of transcriptional regulators, {\it Nature} {\bf 403}, 335 (2000).
\bibitem{ramo}P. R\"am\"o, J. Kesseli and O. Yli-Harja.  Stability of functions in Boolean models of gene regulatory networks {it Chaos} {\bf 15}, 034101 (2005).
\bibitem{shmulevich}I. Shmulevich and S.A. Kauffman.  Activities and sensitivities in Boolean network models, {\it Phys. Rev. Lett.} {\bf 93}, 048701 (2004).
\bibitem{moreira}A.A. Moreira and L.A.N. Amaral.  Canalizing Kauffman networks: nonergodicity and its effect on their critical behavior, {\it Phys. Rev. Lett.} {\bf 94}, 218702 (2005).
\bibitem{derrida}B. Derrida and Y. Pomeau.  Random networks of automata: A simple annealed approximation, {\it Europhys. Lett.} {\bf 1}, 45 (1986).
\bibitem{hwa}N.E. Buchler, U. Gerland and T. Hwa.  On schemes of combinatorial transcription logic, {\it Proc. Natl. Acad. Sci. USA} {\bf 100}, 5136 (2003).
\bibitem{setty}Y. Setty, A.E. Mayo, M.G. Surette and U. Alon.  Detailed map of a cis-regulatory input function, {\it Proc. Natl. Acad. Sci. USA} {\bf 100}, 7702 (2003).
\bibitem{ferrell}J.E. Ferrell, Jr.  Tripping the switch fantastic: how a protein kinase cascade can convert graded inputs into switch-like outputs, {\it Trends in Biochemical Sciences} {\bf 21}, 460 (1996).
\bibitem{kholodenko}B.N. Kholodenko, J.B. Hoek, H.V. Westerhoff and G.C. Brown.  Quantification of information transfer via cellular signal transduction pathways, {\it FEBS Letters} {\bf 414}, 19166 (1997).
\bibitem{hooshangi}S. Hooshangi, S. Thiberge and R. Weiss.  Ultrasensitivity and noise propagation in a synthetic transcriptional cascade, {\it Proc. Natl. Acad. Sci.} {\bf 102}, 3581 (2005).
\bibitem{hermsen}R. Hermsen, S. Tans and P.R. ten Wolde.  Transcriptional regulation by competing transcription factor modules, {\it PLoS Computational Biology} {\bf 2}, 1552 (2006).
\bibitem{buchler1}N.E. Buchler and M. Louis.  Molecular Titration and Ultrasensitivity in Regulatory Networks, {\it J. Molec. Biol.} {\bf 384}, 1106 (2008).
\bibitem{buchler2}N.E. Buchler and F.R. Cross.  Protein sequestration generates a flexible ultrasensitive response in a genetic network, to appear in {\it J. Molec. Biol.} (2009).
\end{thebibliography}
\end{document}